# Ferrimagnetic Order in Tetragonal Antiperovskite Mn$_3$GeN

Shaun O'Donnell[1,2], Corlyn Regier[2], Sharad Mahatara[1], H. Cein Mandujano[3], Efrain E. Rodriguez[3], Danielle R. Yahne[4], Stephan Lany[1], Sage R. Bauers[1], Rebecca W. Smaha[1], James R. Neilson[2]

1. National Renewable Energy Laboratory, Golden, CO
2. Colorado State University, Department of Chemistry, Fort Collins, CO
3. University of Maryland, Department of Chemistry and Biochemistry, College Park, MD
4. Neutron Scattering Division, Oak Ridge National Laboratory, Oak Ridge, 37831, TN

**Abstract**

The crystal and magnetic structures of the nitride antiperovskite Mn$_3$GeN reveals ferrimagnetic order stemming from a distorted kagome-derived lattice of the Mn atoms. Polycrystalline Mn$_3$GeN was synthesized via a solid-state reaction and characterized using neutron powder diffraction, DC magnetometry, and first-principles calculations. Rietveld refinement reveals near-stoichiometric composition (Mn$_3$GeN$_{0.94(1)}$) adopting a tetragonal *I4/mcm* structure at $T$ = 500 K and below, featuring axially distorted and tilted [NMn$_6$] octahedra that result in a buckled Mn kagome lattice. On heating, the tetragonal distortion and octahedral tilt angle decrease continuously before transitioning to the cubic $Pm\bar{3}m$ antiperovskite phase at $T \approx$ 524 K. Neutron diffraction and magnetometry together reveal noncollinear ferrimagnetic ordering. For 30 K ≤ $T$ ≤ 500 K, the magnetic structure is described by a single propagation vector, **k** = (0, 0, 0), with inequivalent Mn1 and Mn2 sublattices that couple antiferromagnetically to yield a net moment. Density functional theory-based calculations show the different local moments originate from the bandwidths associated with the distinct Mn-N bond lengths. The temperature dependence of the sublattice moments indicates a compensation-like crossover between Mn1- and Mn2-derived magnetization near 380 K. These findings uncover a previously unrecognized subtlety in the magnetic and structural behavior of Mn$_3$GeN, highlighting the interplay between structural distortions, magnetic ordering, and electronic structure in kagome-derived antiperovskite materials.

## I. Introduction

Antiperovskite nitrides and carbides exhibit intriguing topological, magnetic, and transport phenomena, typically related to the kagome-derived structure formed by the 3*d* transition metals.[1–4] The kagome lattice is a two-dimensional network of corner-sharing triangles that enforces significant magnetic frustration in the presence of antiferromagnetic spin interactions. The antiperovskite structure with the general composition, $A_3BX$, in which *A* is typically a transition metal, *B* is a metal or main group element, and *X* is a small atom like B, N or C, exhibits 4-intersecting kagome lattices composed of the *A* site cations along the {111} planes. Unless purely ferromagnetic interactions are present, geometric frustration suppresses simple collinear order and instead favors noncollinear magnetic states influenced by small changes to the crystal structure or electronic environment.

Manganese nitride antiperovskites ($Mn_3BN$) have drawn considerable attention for their magnetic and magnetostructural behavior, including topological antiferromagnetism and associated anomalous Hall responses[5–7], frustrated noncollinear antiferromagnetic order[8], magnetically driven negative thermal expansion (NTE)[9–11], and giant barocaloric or magnetocaloric effects.[12] The materials typically exhibit noncollinear antiferromagnetic (AFM) ground states.[8] These magnetic configurations are often described in terms of two closely related AFM configurations ($\Gamma^{4g}$ and $\Gamma^{5g}$) which differ mainly in the orientation of the 120° spin triangles relative to the cubic axes, as recently reviewed.[13] The $\Gamma^{5g}$ state, along with states derived from a mixture with $\Gamma^{4g}$, often leads to chiral spin arrangements.

In some instances, $Mn_3BN$ (e.g. *B* = Cu, Sb, SG: *P4/mmm*) compounds break the degeneracy of the kagome lattice by adopting tetragonal crystal structures.[14,15] Upon ordering, these noncubic compounds adopt more complex ferrimagnetic (FiM) structures with larger magnetic unit cells and small net moments.[8] The combination of small net moments with strong magnetoelastic coupling and field-sensitive magnetostructural transitions makes them candidates for low-energy switching applications.[16,17] The tetragonal distortion within this subset of compounds is quite small as their crystal structures are essentially identical to the cubic phases, but with slight elongations or contractions of the *c* axis. Thus, it is apparent that even minor changes to the Mn sublattice can significantly alter the magnetic landscape of these compounds and lead to alternate complex magnetic structures.

Amongst this large family of compounds, $Mn_3GeN$ is unusual in that its tetragonally distorted structure also exhibits anisotropic Mn-N bonds and out-of-phase [$NMn_6$] octahedral tilting. At room temperature, $Mn_3GeN$ is a tetragonally distorted perovskite (space group I*4/mcm*) with rotated [$NMn_6$] octahedra.[18,19] These distortions lead to a buckling of the kagome layers, a feature absent from any of the other known compounds in this family. This buckling breaks the in-plane symmetry of the kagome lattice and allows for in-plane Dzyaloshinskii–Moriya (DM) interactions.

However, the $Mn_3GeN$ magnetic ground state and its relationship to its crystal structure remained ambiguous. One prior study reports that for $T < 380$ K, $Mn_3GeN$ has a ferromagnetic (FM) ground state, then undergoes a magnetic transition into an AFM state and finally undergoes a coupled structural and magnetic transition to a paramagnetic (PM) cubic $Pm\bar{3}m$ phase at 600 K.[18] A second study of $Mn_3Cu_{1-x}Ge_xN$ solid solutions suggests that $Mn_3GeN$ is FM up to $T = 480$ K, then transitions to a cubic AFM structure, and finally thermally disorders at $T = 510$ K.[8] A third report states that $Mn_3GeN$ has a single coupled structural and magnetic transition, where the tetragonal phase has a noncollinear FiM ground state derived from the $\Gamma^{5g}$ magnetic structure which becomes PM and cubic at elevated temperature.[13] An additional open question is with respect to the nitrogen stoichiometry in $Mn_3GeN$, since when it is explicitly mentioned, the nitrogen stoichiometry is always the sub-stoichiometric $Mn_3GeN_{0.75}$. It is known that the magnetic properties of several other nitride antiperovskites are intimately tied to nitrogen composition. For example, the Néel temperature of $Mn_3GaN$ increases by more than 75 K for the sub-stoichiometric $Mn_3GaN_{0.83}$, and in the extreme case of complete denitridation, $Mn_3Ga$ becomes FM.[20,21]

However, determination of the nitrogen stoichiometry is often challenging in these materials due to technique-based challenges and the prevalent equilibrium with $N_2(g)$.

In this contribution, we report the powder synthesis of $Mn_3GeN$ and use temperature-dependent neutron powder diffraction (NPD) to determine the stoichiometry, structure, and magnetic configuration. Analysis of the NPD data reveals a stoichiometry of $Mn_3GeN_{0.94(1)}$. From 30 K ≤ $T$ ≤ 500 K, the magnetic structure at zero field is ferrimagnetic, in agreement with the DC magnetization. We identify a tetragonal to cubic structural transition at $T$ = 524 K, commensurate with a loss of magnetic order and magnetization. Density functional theory-based calculations provide insight into the local magnetization through the distinct bandwidths of each Mn environment. Further, DFT–Monte Carlo calculations confirm a FiM ground state, with the AFM state higher in energy by 13 meV/Mn and the FM state strongly unstable (128 meV/Mn). Together, these results establish $Mn_3GeN$ as a structurally and magnetically distinct member of the Mn nitride antiperovskites, in which a buckled kagome lattice leads to a unique ferrimagnetic structure not found in any other compound in this family..

## II. Methods

**A. Synthesis.** The synthesis of $Mn_3GeN$ powders was adapted from prior literature procedures. First, the $Mn_2N$ precursor was prepared by heating Mn powder (99.95%, Alfa Aesar) in a tube furnace under flowing $N_2$ (~50 sccm) for 60 h at 650°C. Next, $Mn_2N$ and Ge powder (≥99.999%, MSE Supplies) were ground together in a 3:2 molar ratio using an agate mortar and pestle for ca. 20 min to obtain a homogenous mixture. The mixed precursor powder was pressed into a pellet and then loaded into a tantalum crucible with a loosely fit lid. The crucible was subsequently sealed into a fused silica ampoule under vacuum (~30 mTorr) and heated in a box furnace at 600°C for 168 h. After heating, the furnace cooled naturally to room temperature. The resulting pellet was ground to obtain the final product. All processing and handling of powders was performed inside an Ar-filled glovebox ($H_2O$ and $O_2$ <0.1ppm).

**B. Characterization.** Neutron powder diffraction (NPD) data were collected using beamline HB-2A at the High Flux Isotope Reaction at Oak Ridge National Laboratory. Data were collected at temperatures of 30 K, 300 K, 400 K, 500 K, 580 K, 600 K, and 700 K. A wavelength of 2.409897 Å was used at all temperatures, while additional datasets using a wavelength of 1.537612 Å were collected at 30 K, 300 K, and 700 K. Rietveld refinements of the nuclear and magnetic structures were carried out using the FullProf program.[22,23] Refinement of the nuclear structure was initialized using a previously reported structure for $Mn_3GeN_{0.75}$ in the *I4/mcm* space group.[19] After refining the NPD data against the nuclear structure, the remaining unindexed intensities were used to determine the possible *k*-vectors of the magnetic structure. Possible magnetic *k*-vectors were identified using the k-search program within FullProf. For 30 ≤ *T* ≤ 500 K, k = (0,0,0) indexed all reflections. For *T* = 30 K, additional reflections were observed from the magnetic structure of MnO.[24] For *T* ≥ 580 K, all reflections were indexed with the $Pm\bar{3}m$ space group corresponding to the nuclear structure of many Mn-based antiperovskite nitrides.[8] Representational analysis using SARAh was performed to generate the candidate irreducible representations and basis vectors; all basis vectors were converted to real vectors.[25] Where applicable, co-refinements using both neutron wavelengths were used to determine the nuclear and magnetic structures. A small MnO impurity was present in the powder sample, and its nuclear and magnetic structures were included in the refinement. MnO magnetic orders below *T* = 115 K. For the 30 K data, the MnO magnetic structure was modeled using the monoclinic magnetic space group, $C_c2/c$.[24]

Temperature- and field-dependent magnetization was measured using a Quantum Design Dynacool Physical Properties Measurement System (PPMS) using the ACMS-II option. Powder samples were embedded in superglue prior to measurement to prevent realignment in the magnetic field. Isothermal DC magnetization as a function of applied field (−14 T ≥ $\mu_0$H ≥ 14 T) was measured from 2 K to 350 K. Variable temperature DC susceptibility data were measured from 2 K to 300 K under static applied fields ranging from 0.01 T to 1 T. A high temperature dataset from 300 K to 700 K was measured using a Quantum Design MPMS3 with the high temperature furnace using an applied field of $\mu_0H$ = 0.1 T. The sample preparation for the high temperature data was as follows. A pressed pellet of the precursors was annealed following the synthesis procedure detailed above. A piece of this sintered pellet was then attached to the MPMS heating rod using Zircar® cement. However, because Mn$_3$GeN reacts with water, the bottom portion of the Mn$_3$GeN pellet was partially decomposed due to the water-based Zircar® cement. As such, this high temperature data was only used to determine the paramagnetic transition temperature. A plot of the full 0.1 T data from 2 K to 700 K was generated by merging these datasets. Due to the partial decomposition of Mn$_3$GeN in the high temperature data, its susceptibility did not overlap with the low temperature data around the 300 K region. As such, an empirical constant scaling factor of 2.12 was applied to the high temperature susceptibility data.

**C. Calculations.** The magnetic ground state of Mn$_3$GeN was explored via first-principles Monte Carlo (MC) simulations, using Density Function Theory (DFT) based total energies as the acceptance criterion on a 2x2x2 supercell of the primitive cell of the canonical $Pm\bar{3}m$ antiperovskite structure (40 atoms; 24 Mn), as to prevent any bias from symmetry-lowering distortions and their coupling to magnetic order.[26–28] Simulations were initialized with random moments and run at 1000, 500, and 100 K, typically converging within 300 MC steps at ≤ 1000 K. Both antiferromagnetic (AFM) and ferrimagnetic (FiM, 2:1 Mn spin-up:spin-down) seeds were considered, and in each case, we were able to converge the calculations to the magnetic state configuration targeted by the initialization of the MC trial, indicating that both AFM and FiM are locally stable solutions.[29–31] At least three random magnetic seeds were evaluated at each temperature.

All candidate structures were relaxed (atom positions and unit cells) using a spin-polarized GGA+$U$ functional (PBE, $U$–$J$=1 eV for Mn 3$d$) in VASP with a 4x4x4 $k$-point mesh, a 350 eV cutoff, and convergence criteria of 10$^{-5}$ eV for total energy and 0.02 eV Å$^{-1}$ for forces. The Hubbard $U$ correction was used to stabilize the convergence of each magnetic solution. Here,

PAW potentials from VASP 4.6 were applied, including a soft N potential.[32] Minimum-energy structures were further relaxed with GGA functional (without U) using a 6x6x6 k-point mesh and a 0.01 eV Å$^{-1}$ force threshold.[33] Partial density of states (PDOS) for Mn, Ge, and N were computed using integration spheres of radius 1 Å and Gaussian broadening with σ = 0.05 eV.

## III. Results and Discussion

### A. Nuclear Structure and Structural Phase Transition

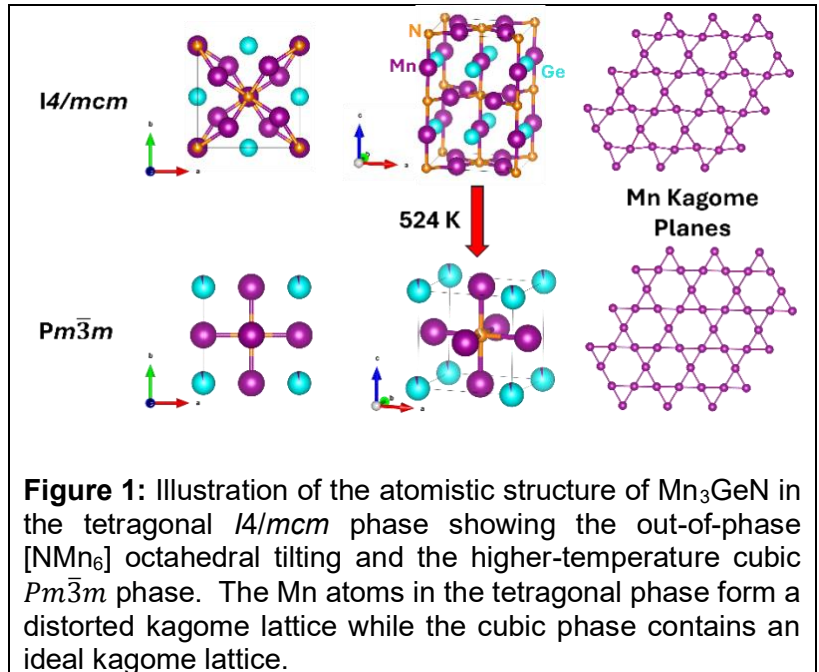

**Figure 1:** Illustration of the atomistic structure of Mn$_3$GeN in the tetragonal *I4/mcm* phase showing the out-of-phase [NMn$_6$] octahedral tilting and the higher-temperature cubic $Pm\bar{3}m$ phase. The Mn atoms in the tetragonal phase form a distorted kagome lattice while the cubic phase contains an ideal kagome lattice.

The nuclear structures at temperatures ranging from 30-700K were determined via Rietveld refinement of NPD data. Accurate determination of the nuclear structure required co-determination of the magnetic structure due to the presence of many overlapping reflections. The NPD patterns between 30 K and 500 K were consistent with the prior reported nuclear structure for Mn$_3$GeN$_{0.75}$ in space group #140 (*I4/mcm*).[19] The reflections in the NPD from 580 ≤ T ≤ 700K are indexed with space group #221 ($Pm\bar{3}m$). The tetragonal to cubic transition temperature, $T \cong 524$ K, is inferred from magnetometry results discussed later. A small MnO impurity of ~6 mol% was identified and is a common impurity in many manganese antiperovskite powders.

The tetragonal crystal structure exists with axial distortions and tilting of the octahedra in the antiperovskite structure. The crystal structures of the lower and higher temperature phases are shown in **Figure 1**. The lower-temperature tetragonal structure contains two crystallographic Mn sites at 8h (Mn1) and 4a (Mn2), and one Ge and N site at 4b and 4c, respectively. The structure consists of corner-sharing [NMn$_6$] octahedra that are rotated within the ab-plane, and out-of-phase along the c-axis ($a^0a^0c^-$ in Glazer notation). These rotations lead to a distortion in the Mn kagome planes, as illustrated in **Figure 1**. Within the triangles which make up the kagome lattice, two of the corners are Mn1 and 1 corner is Mn2, thus breaking the equilateral symmetry. The octahedral rotation causes one of the Mn1 sites to be pushed above the plane, and the other below the plane, resulting in the formation of a buckled layer. The [NMn$_6$] octahedra are also tetragonally distorted, as the axial Mn-N bonds are longer than the equational ones (e.g. 2.0312(1) Å and 1.9608(7) Å at 300 K).

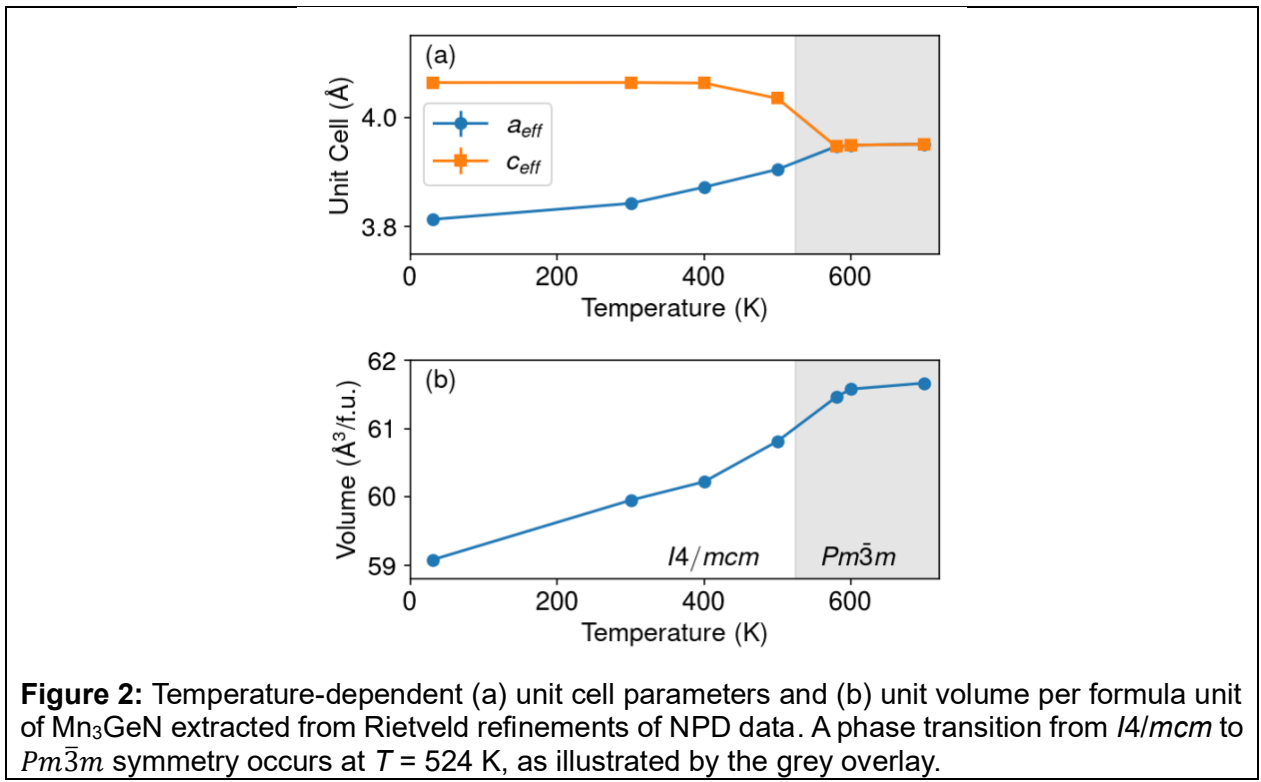

**Figure 2:** Temperature-dependent (a) unit cell parameters and (b) unit volume per formula unit of Mn$_3$GeN extracted from Rietveld refinements of NPD data. A phase transition from $I4/mcm$ to $Pm\bar{3}m$ symmetry occurs at $T$ = 524 K, as illustrated by the grey overlay.

The tetragonal distortion disappears on warming. As shown in **Figure 2(a)**, the effective cubic lattice parameter, $a_{eff} = a_{tet}/\sqrt{2}$, gradually increases with temperature up to 500 K. The effective cubic lattice parameter, $c_{eff} = c_{tet}/2$, is nearly constant before decreasing from 400 K to 580 K. The coalescence of the $a_{eff} = c_{eff}$ coincides with the phase transition to cubic symmetry. Throughout this change, the unit cell volume continually increases (**Figure 2(b)**), showing an overall positive coefficient of thermal expansion. This gradual transition towards the cubic structure is similarly reflected in the degree of tetragonality and octahedral tilting. Shown in **Figure 3(a)**, the $a_{eff}/c_{eff}$ ratio increases towards 1 from 30 K to 580 K; this trend follows the untilting of the [NMn$_6$] octahedra, defined by the Mn-N-Mn bond angle in the tetragonal *ab* plane.

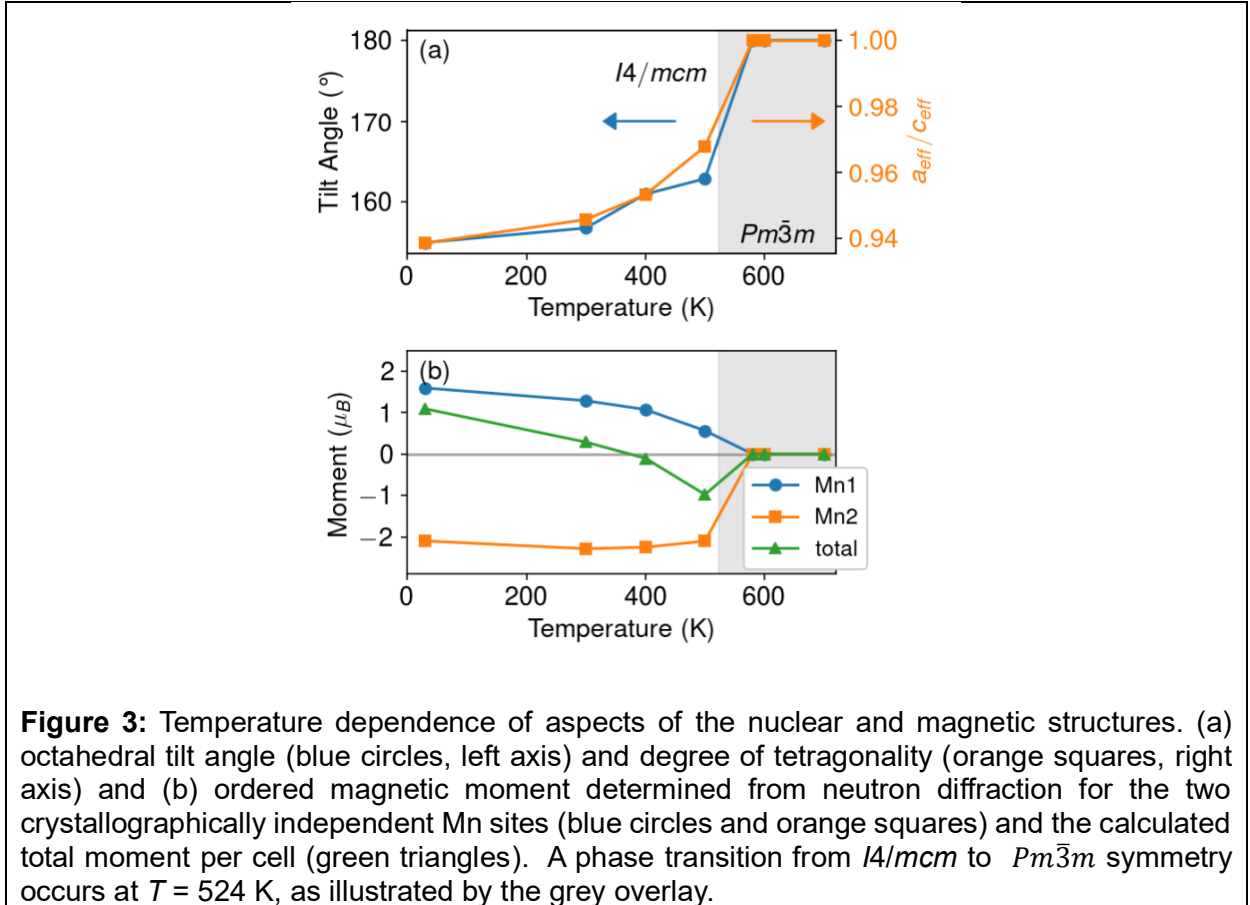

**Figure 3:** Temperature dependence of aspects of the nuclear and magnetic structures. (a) octahedral tilt angle (blue circles, left axis) and degree of tetragonality (orange squares, right axis) and (b) ordered magnetic moment determined from neutron diffraction for the two crystallographically independent Mn sites (blue circles and orange squares) and the calculated total moment per cell (green triangles). A phase transition from *I*4/*mcm* to $Pm\bar{3}m$ symmetry occurs at *T* = 524 K, as illustrated by the grey overlay.

Above 524 K the nuclear structure transitions to the canonical cubic antiperovskite structure. Here, the Mn, Ge, and N occupy the 3*c*, 1*a*, and 1*b* sites of space group $Pm\bar{3}m$, respectively. Additionally, there is no longer any rotation of the [NMn$_6$] octahedra. Refinement of the site occupancies indicates a small amount of Mn/Ge site disorder, with ~7% Mn and Ge disorder. The low temperature data in the tetragonal polymorph does not show signs of site disorder, suggesting that the small Mn-Ge site mixing is induced at higher temperatures and potentially driven by the structural phase transition. The lattice parameter shows a gradual thermal expansion with increasing temperature.

**Table I:** Structure parameters of $I4/mcm$ $Mn_3GeN$ with the fractional coordinates Mn1 = x, x+1/2, 0; Mn2 = 0 0 ¼, Ge1 = 0, 0.5, 0.25, and N1 = 0, 0, 0, as well as the magnetic basis vector coefficients for the basis vectors defined in **Table III**.

| Temperature (K) | 30 | 300 | 400 | 500 |
|---|---|---|---|---|
| a (Å) | 5.3928(6) | 5.432867(0) | 5.45534(6) | 5.50108(7) |
| c (Å) | 8.126(1) | 8.125064(0) | 8.0941(2) | 8.0383(2) |
| Mn1 x | 0.194(3) | 0.1987(4) | 0.2082(8) | 0.2124(8) |
| N1 occ. | 0.94 | 0.94(1) | 0.94 | 0.94 |
| Mn1 C1 | 0.72(1) | 0.64(2) | 0.52(2) | 0.22(2) |
| Mn1 C2 | 0.1(1) | –0.06(3) | -0.12(3) | -0.17(3) |
| Mn2 C3 | -0.53(3) | –0.57(3) | -0.56(2) | -0.53(2) |

Refinement of the nitrogen occupancy reveals nearly stoichiometric $Mn_3GeN$, with a slight reduction of nitrogen content at elevated temperature. Free refinement of the nitrogen occupancy carried out for data collected at 300 K and 700 K (temperatures which contained diffraction patterns measured with both neutron wavelengths) resulted in 0.94(1) and 0.89(1), respectively. The overlapping magnetic and nuclear contributions to some of the Bragg peaks partially convolve the refined magnetic moment with nitrogen occupancy. The magnetic contribution is eliminated with the 700 K data, but since elevated temperature leads to nitrogen loss in other Mn antiperovskite nitrides, it is possible that some nitrogen was lost when the sample was at 700 K, resulting in a slightly lower nitrogen value.[20] Refinement of the nitrogen occupancy in the data collected at 580 K (with neutrons of only 2.41 Å wavelength) results in a value of 0.933(7), almost identical to the 300 K value. Altogether, this suggests that the as-synthesized nitrogen stoichiometry is ~0.94 and is much greater than the previously reported value of 0.75. For simplicity, we refer to the compound as Mn$_3$GeN.

**B. Magnetic Structure and Phase Transition.**

The magnetic structure of Mn$_3$GeN is a non-collinear ferrimagnet described by a single propagation vector, ***k*** = (0, 0, 0). Representational analysis yields a decomposition of the magnetic representation into a total of 9 possible irreducible representations for Mn1 ($\Gamma_{\text{mag}} = \Gamma_2 + \Gamma_3 + \Gamma_4 + \Gamma_5 + \Gamma_6 + \Gamma_8 + 2\Gamma_9 + \Gamma_{10}$) and 4 possible irreducible representations for Mn2 ($\Gamma_{\text{mag}} = \Gamma_3 + \Gamma_4 + \Gamma_9 + \Gamma_{10}$). We constrained each site to a single irreducible representation without a distinction of moment directions within the *ab* plane (due to powder and domain averaging), thus limiting to only 60 possible permutations. We tested each by Rietveld refinement and one model best describes the data, with Rietveld refinement yielding $R_{mag}$ = 6.51% and 7.31% for the ~2.41 Å and ~1.54 Å wavelength data, respectively at 30 K. Several other combinations produced nearly as good solutions and were all qualitatively similar to the solution presented here but with slightly different moment magnitudes. As such, we discuss the structural trends that are independent of the exact choice of model. The calculated patterns and observed diffraction patterns are provided as Supplementary Material.

The NPD are best described by a ferrimagnetic configuration with different moment values for Mn1 and Mn2 spanned by irreducible representation $\Gamma_9$ for both sites for 30 K ≤ *T* ≤ 500 K. This noncollinear structure is illustrated in **Figure 4**, and the set of basis vectors is tabulated in **Table II**. For Mn1, basis vector $\Psi_1$ describes a ferromagnetic component along the *a*-axis while basis vector $\Psi_2$ presents an antiferromagnetic component along *b*. For Mn2, basis vector $\Psi_3$ is ferromagnetic along the *a*-axis. The resulting refined magnetic structure is ferrimagnetic and derived from A-type antiferromagnetic ordering where planes of Mn1 and Mn2 moments are coupled antiferromagnetically to each other along the *a*-axis. The Mn1 vector also contains a small *b*-axis component that is AFM between neighboring Mn1 atoms (**Figure 4(b)**) such that a net moment only persists along *a*. When viewed from the perspective of the kagome layers, this magnetic structure forms AFM stripes between neighboring Mn1 and Mn2 (**Figure 4(c)**). The magnitude of the individual moments at 30 K are equal to 1.60 $\mu_B$ for Mn1 and −2.10 $\mu_B$ for Mn2, resulting in a net moment of 1.10 $\mu_B$ f.u.$^{-1}$. These values agree with the DFT-MC calculations and the DC magnetometry reported below.

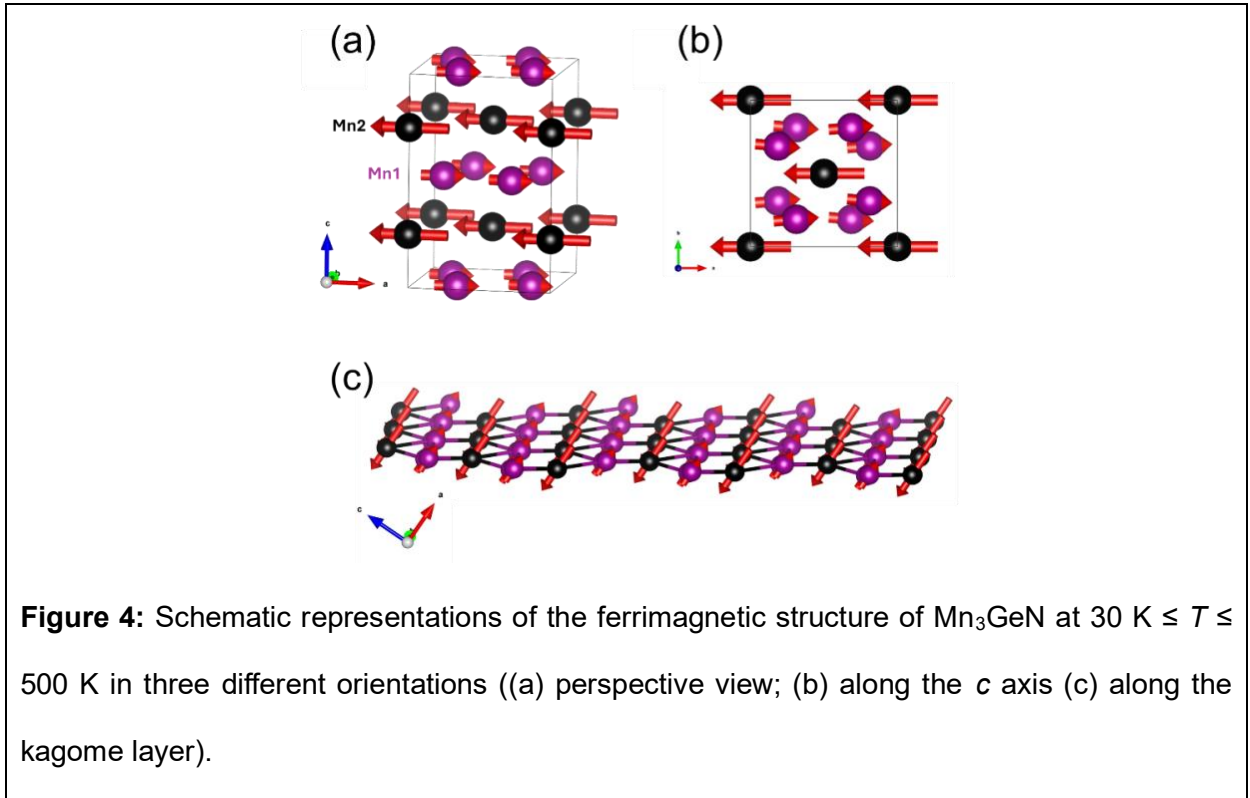

**Figure 4:** Schematic representations of the ferrimagnetic structure of Mn$_3$GeN at 30 K ≤ *T* ≤ 500 K in three different orientations ((a) perspective view; (b) along the *c* axis (c) along the kagome layer).

This ferrimagnetic structure is retained on warming with gradual reduction in the ordered moment. Rietveld refinement with the same magnetic structure yields similarly good fits for the data collected at *T* = 300 K ($R_{mag}$ = 5.96%, 8.07%), *T* = 400 K ($R_{mag}$ = 5.20%), and 500 K ($R_{mag}$ = 8.35%). The temperature dependence of the site-decomposed moments and total ordered moments per unit cell is illustrated **Figure 3(b).** The magnitudes of the Mn1 and Mn2 moments both decrease with increasing temperatures but with different slopes: the Mn1 substructure thermally disorders at a lower temperature than the Mn2 substructure, consistent with the higher moment of the Mn2 sites. Therefore, the net moment from 300 K to 400 K flips sign and decreases in magnitude to –0.10 $\mu_B$ f.u.$^{-1}$ at 400 K and then increases to –0.98 $\mu_B$ f.u.$^{-1}$ at 500 K before thermally disordering at 580 K.

**Table II:** Real components of the basis vectors in the $\Gamma_9$ irreducible representation for space group *I4/mcm* used to fit the k = (0, 0, 0) magnetic structure. The decomposition of the magnetic representation for the Mn1 site are atom: 1: (x, x+1/2, 0); 2: (x, 1/2–x, 1/2); 3: (–x, 1/2–x, 0.5); 4: (–x, x+1/2, 0). For Mn2 site, atom: 5: (0, 0, ¼); 6: (0, 0, ¾).

| IR | BV | Atom | $m_x$ | $m_y$ | $m_z$ |
|---|---|---|---|---|---|
| $\Gamma_9$ (Mn1) | $\psi_1$ (C1) | 1 | 2 | 0 | 0 |
| | | 2 | 2 | 0 | 0 |
| | | 3 | 2 | 0 | 0 |
| | | 4 | 2 | 0 | 0 |
| | $\psi_2$ (C2) | 1 | 0 | 2 | 0 |
| | | 2 | 0 | –2 | 0 |
| | | 3 | 0 | –2 | 0 |
| | | 4 | 0 | 2 | 0 |
| $\Gamma_9$ (Mn2) | $\psi_3$ (C3) | 5 | 4 | 0 | 0 |
| | | 6 | 4 | 0 | 0 |

### C. Density Functional Theory Calculations

DFT-MC structural relaxation indicates that magnetic ordering strongly influences the crystal structure of Mn$_3$GeN. All configurations exhibit a symmetry-lowering distortion. As the system evolves from the FM to the FiM state, the in-plane lattice parameter increases by 1.4%, while the *c*-axis lattice parameter decreases by 1.6%, leading to an overall expansion of the unit-cell volume (**Table III**). The increased volume in the FiM and AFM states reflects the lattice response to the spin-up/spin-down arrangement of Mn moments and the associated symmetry lowering. The FiM configuration is the ground state, with the FM state lying significantly higher in energy by 128 meV/Mn (see **Table III**). In addition, the mean energy of the relaxed FiM state is 13 meV/Mn lower than that of the lowest-energy AFM configurations. Thus, the DFT calculations support the experimental observation of a FiM ground state.

**Table III:** DFT-relaxed (GGA without U) lattice parameters of $Mn_3GeN$ in a 40-atom supercell ($\alpha=\beta=\gamma=90°$) and the relative energies of FiM, FM, and AFM magnetic configurations. ΔE is relative to the FiM ground state.

| DFT results | FM | FiM | AFM |
|---|---|---|---|
| $a, b$ (Å) | 7.399 | 7.498 | 7.476, 7.531 |
| $c$ (Å) | 8.200 | 8.065 | 8.034 |
| $a_{eff}/c_{eff}$ | 0.902 | 0.926 | 0.930-0.937 |
| N-Mn-N angle (°) | 150.6 | 152.4 | 146.2-152.1 |
| Volume (Å$^3$) | 448.860 | 453.441 | 452.367 |
| $\Delta E$ (meV/Mn) | 128 | 0 | 13 |

The DFT-MC calculations for a FM ground state in a 40-atom unit cell yield two symmetry-inequivalent Mn sites with magnetic moments of 1.7 $\mu_B$ (Mn1) and 2.6 $\mu_B$ (Mn2). In contrast, the FiM ground state obtained from DFT–MC calculations show a reduced distinction between Mn1 and Mn2 moments, with values of 2.1 $\mu_B$ and −2.5 $\mu_B$, respectively, resulting in a net magnetization of ~1.7 $\mu_B$ f.u.$^{-1}$. The Mn1–Mn2 moment difference is underestimated relative to experiment, likely reflecting known limitations of GGA. AFM configurations generated within the DFT–MC framework yield finite net magnetizations, with the lowest energy configuration yielding −0.45 $\mu_B$ f.u.$^{-1}$. This indicates that a fully compensated AFM state is not stabilized and that the system retains a FiM-like behavior.

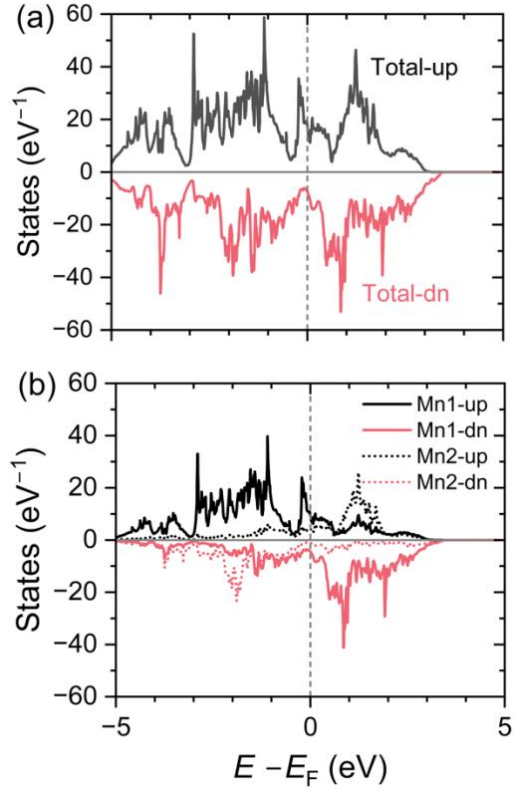

**Figure 5:** (a) Total DOS of FiM $Mn_3GeN$ and (b) site-resolved PDOS of Mn1 (solid) and Mn2 (dotted), computed using standard GGA; the Fermi level is set to zero energy.

The total density of states (DOS) for the FiM configuration exhibits an asymmetric distribution of spin-up and spin-down channels, consistent with the net magnetization (**Figure 5(a)**). Site-projected moments show Mn1 (2.1 $\mu_B$) is slightly smaller than Mn2 (2.5 $\mu_B$), reflecting differences in their local bonding environments. As seen in the PDOS (**Figure 5(b)**), Mn1 displays a broader bandwidth, whereas Mn2 exhibits a narrower bandwidth, consistent with more localization occurring from the longer Mn–N bonds of the Mn2 site rather than the distorted N-Mn-N bond angles observed in the Mn1 sites. These results reveal that the variation in local moments arises from the distinct Mn–N bond lengths and associated bandwidths.

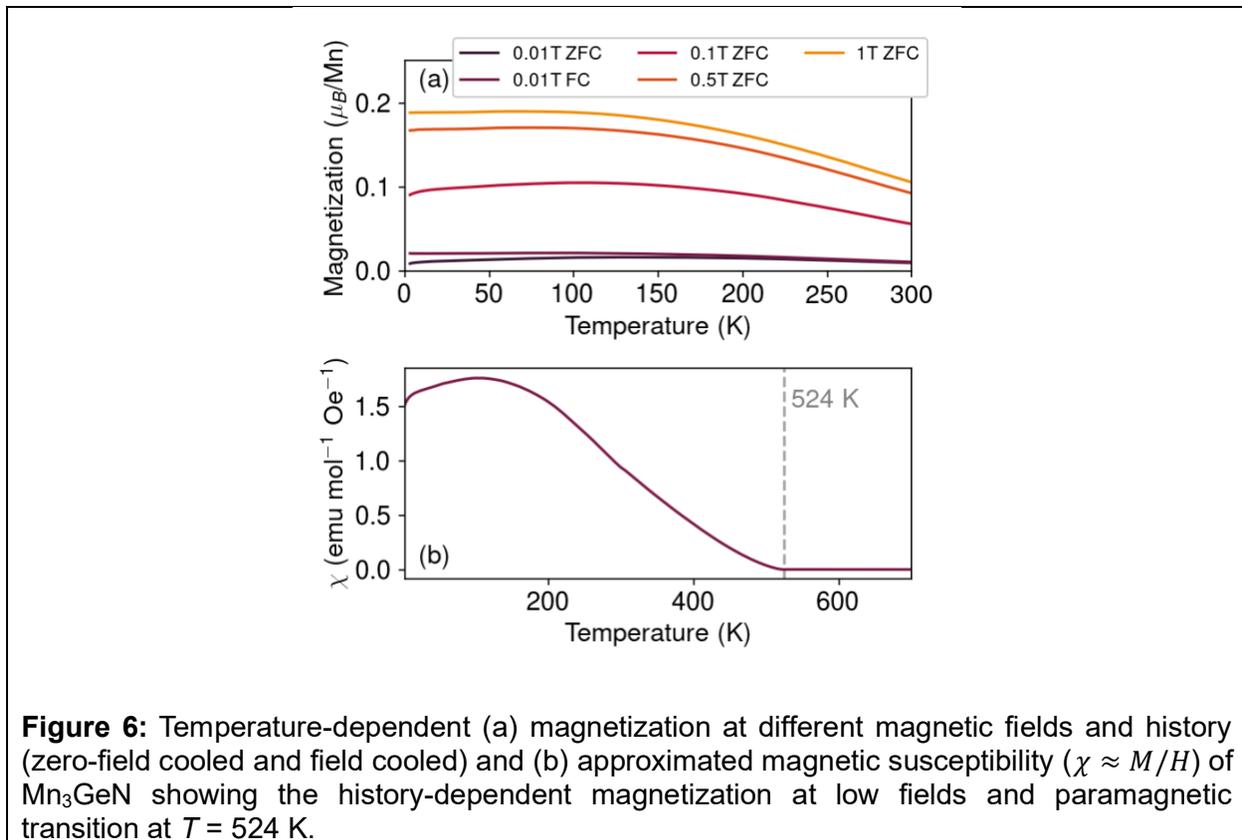

**Figure 6:** Temperature-dependent (a) magnetization at different magnetic fields and history (zero-field cooled and field cooled) and (b) approximated magnetic susceptibility ($\chi \approx M/H$) of Mn$_3$GeN showing the history-dependent magnetization at low fields and paramagnetic transition at $T$ = 524 K.

### D. Magnetic Properties.

The temperature-dependent magnetization reveals a magnetic ordering temperature around 524 K but no other major transitions at lower temperatures. Between $2 \leq T \leq 300$ K, the magnetization shown in **Figure 6(a)** is consistent with FiM behavior and in agreement with the magnetic structure solutions in this temperature range. A plateau in the magnetization below $T \sim$ 100 K is consistent with saturation of the magnetic order parameter (**Figure 3(b)**). At low field ($\mu_0 H$ = 0.01 T), there is a slight history dependence below 100 K (ZFC/FC splitting), possibly due to the formation of ferrimagnetic domains within the sample.. The susceptibility (approximated by $\chi \approx M/H$, $\mu_0 H$ = 0.01 T) measured up to 700 K illustrates the decreasing magnetic order parameter (**Figure 6(b)**) to the paramagnetic transition at $T \sim$ 524 K. This is significantly lower than 600 K, which was the value previously reported.[18] Additionally, this previous work reported a

FM to AFM transition at 380 K, but we see no indication of this transition in our magnetization data nor in the refined magnetic structures. These discrepancies could be tied to the differences in nitrogen stoichiometry between our materials, which is known to influence the magnetic properties of other nitride antiperovskites.

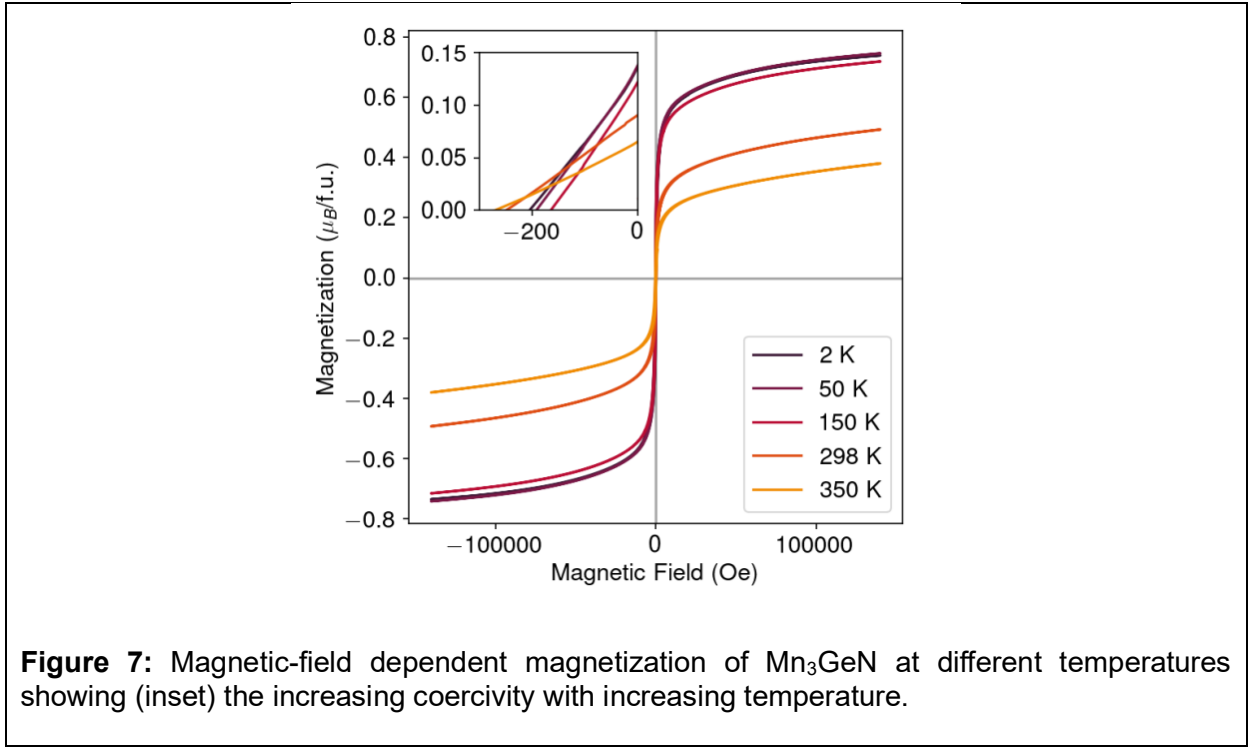

**Figure 7:** Magnetic-field dependent magnetization of $Mn_3GeN$ at different temperatures showing (inset) the increasing coercivity with increasing temperature.

Field dependent, isothermal magnetization data reveals soft ferrimagnetic behavior consistent with the different ordering behavior of two magnetic sublattices. Shown in **Figure 7**, at $T$ = 2 K, the saturation magnetization is relatively low, 0.74 $\mu_B$/f.u., consistent with the ferrimagnetic order. This does not quantitatively agree with the total ordered moment determined from neutron diffraction, which likely reflects the incomplete saturation at $\mu_0 H$ = ±14 T. The coercivity is quite small, only on the order of ~100-300 Oe. Interestingly, the coercivity initially decreases with temperature before increasing and then reaching a maximum (270 Oe) at 350 K, the highest temperature measured. This occurs when a ferrimagnet approaches its compensation point of the two different magnetic sublattices ($H_c \approx 2K/\mu_0 M$);[34] this is consistent with the temperature dependence of the total ordered moment per unit cell observed by neutron diffraction from 30 K to 500 K (**Figure 3(b)**), in that a compensation point from Mn1-dominant magnetic order to Mn2-dominant magnetic order switches over between 300 K and 500 K, likely around 380 K. Together, the magnetization reflects the observed ferrimagnetic order.

## IV. Conclusions

Neutron powder diffraction and magnetometry together show that $Mn_3GeN$ is a ferrimagnet in its ground state. Structurally, $Mn_3GeN$ adopts a tetragonal *I4/mcm* crystal structure at lower temperatures with out-of-phase tilting of corner-sharing [$NMn_6$] octahedra coupled with an axial distortion ($a^0a^0c^-$). These distortions break the degeneracy of the kagome lattice of Mn atoms described in the cubic phase. On heating, the tetragonal distortion and octahedral tilt angle gradually decrease, and the structure undergoes a transition to a cubic antiperovskite phase at $T \approx 524$ K, well below the previously reported structural/magnetic transition temperature. $Mn_3GeN$ is a noncollinear ferrimagnet over the entire ordered temperature range. For 30 K ≤ T ≤ 500 K, the magnetic structure is described by a single $k$ = (0,0,0) propagation vector with inequivalent Mn1 and Mn2 sublattices that couple antiferromagnetically, yielding a small net moment per formula unit. Collinear DFT–MC calculations identify the FiM ground state, capturing distinct local moments at Mn1 (2.1 $\mu_B$) and Mn2 (−2.5 $\mu_B$) sites, though the moment difference is underestimated relative to experiment. Projected DOS suggests that this difference likely arises from the varying bandwidths most closely associated with the distinct Mn–N bond lengths rather than symmetry lowering distortions. Put together, these data show the rich structural and magnetic features present in $Mn_3GeN$ highlighting the manifestation of frustrated interactions.

## V. Acknowledgements

This work was authored in part by the National Laboratory of the Rockies for the U.S. Department of Energy (DOE), operated under Contract No. DE-AC36-08GO28308. Funding provided by the

U.S. Department of Energy, Office of Science, Basic Energy Sciences, Division of Materials Science, through the Office of Science Funding Opportunity Announcement (FOA) Number DE-FOA-0002676: Chemical and Materials Sciences to Advance Clean-Energy Technologies and Transform Manufacturing. The views expressed in the article do not necessarily represent the views of the DOE or the U.S. Government. A portion of this research used resources at the High Flux Isotope Reactor, a DOE Office of Science User Facility operated by the Oak Ridge National Laboratory. The beam time was allocated to HB-2A on proposal numbers IPTS-32084.1 and IPTS-34370.1. The authors thank the Analytical Resources Core (ARC) at Colorado State University (CSU) for training and access to instruments (RRID: SCR_021758).

## VI. References

(1) Wang, Q.; Lei, H.; Qi, Y.; Felser, C. Topological Quantum Materials with Kagome Lattice. *Acc. Mater. Res.* **2024**, *5* (7), 786–796. https://doi.org/10.1021/accountsmr.3c00291.
(2) Nayak, A. K.; Fischer, J. E.; Sun, Y.; Yan, B.; Karel, J.; Komarek, A. C.; Shekhar, C.; Kumar, N.; Schnelle, W.; Kübler, J.; Felser, C.; Parkin, S. S. P. Large Anomalous Hall Effect Driven by a Nonvanishing Berry Curvature in the Noncolinear Antiferromagnet $Mn_3$ Ge. *Sci. Adv.* **2016**, *2* (4), e1501870. https://doi.org/10.1126/sciadv.1501870.
(3) Nakatsuji, S.; Kiyohara, N.; Higo, T. Large Anomalous Hall Effect in a Non-Collinear Antiferromagnet at Room Temperature. *Nature* **2015**, *527* (7577), 212–215. https://doi.org/10.1038/nature15723.
(4) Tanaka, M.; Fujishiro, Y.; Mogi, M.; Kaneko, Y.; Yokosawa, T.; Kanazawa, N.; Minami, S.; Koretsune, T.; Arita, R.; Tarucha, S.; Yamamoto, M.; Tokura, Y. Topological Kagome Magnet $Co_3$ $Sn_2$ $S_2$ Thin Flakes with High Electron Mobility and Large Anomalous Hall Effect. *Nano Lett.* **2020**, *20* (10), 7476–7481. https://doi.org/10.1021/acs.nanolett.0c02962.
(5) You, Y.; Bai, H.; Chen, X.; Zhou, Y.; Zhou, X.; Pan, F.; Song, C. Room Temperature Anomalous Hall Effect in Antiferromagnetic Mn3SnN Films. *Appl. Phys. Lett.* **2020**, *117* (22), 222404. https://doi.org/10.1063/5.0032106.
(6) Torres-Amaris, D.; Bautista-Hernandez, A.; González-Hernández, R.; Romero, A. H.; Garcia-Castro, A. C. Anomalous Hall Conductivity Control in $Mn_3$ NiN Antiperovskite by Epitaxial Strain along the Kagome Plane. *Phys. Rev. B* **2022**, *106* (19), 195113. https://doi.org/10.1103/PhysRevB.106.195113.
(7) Zhao, K.; Hajiri, T.; Chen, H.; Miki, R.; Asano, H.; Gegenwart, P. Anomalous Hall Effect in the Noncollinear Antiferromagnetic Antiperovskite $Mn_3$ $Ni_{1-x}$ $Cu_x$ N. *Phys. Rev. B* **2019**, *100* (4), 045109. https://doi.org/10.1103/PhysRevB.100.045109.
(8) Fruchart, D.; F. Bertaut, E. Magnetic Studies of the Metallic Perovskite-Type Compounds of Manganese. *J. Phys. Soc. Jpn.* **1978**, *44* (3), 781–791. https://doi.org/10.1143/JPSJ.44.781.
(9) Takenaka, K.; Takagi, H. Giant Negative Thermal Expansion in Ge-Doped Anti-Perovskite Manganese Nitrides. *Appl. Phys. Lett.* **2005**, *87* (26), 261902. https://doi.org/10.1063/1.2147726.

(10) Takenaka, K.; Asano, K.; Misawa, M.; Takagi, H. Negative Thermal Expansion in Ge-Free Antiperovskite Manganese Nitrides: Tin-Doping Effect. *Appl. Phys. Lett.* **2008**, *92* (1), 011927. https://doi.org/10.1063/1.2831715.

(11) Huang, R.; Li, L.; Cai, F.; Xu, X.; Qian, L. Low-Temperature Negative Thermal Expansion of the Antiperovskite Manganese Nitride Mn3CuN Codoped with Ge and Si. *Appl. Phys. Lett.* **2008**, *93* (8), 081902. https://doi.org/10.1063/1.2970998.

(12) Matsunami, D.; Fujita, A.; Takenaka, K.; Kano, M. Giant Barocaloric Effect Enhanced by the Frustration of the Antiferromagnetic Phase in Mn3GaN. *Nat. Mater.* **2015**, *14* (1), 73–78. https://doi.org/10.1038/nmat4117.

(13) Rimmler, B. H.; Pal, B.; Parkin, S. S. P. Non-Collinear Antiferromagnetic Spintronics. *Nat. Rev. Mater.* **2024**, *10* (2), 109–127. https://doi.org/10.1038/s41578-024-00706-w.

(14) Iikubo, S.; Kodama, K.; Takenaka, K.; Takagi, H.; Shamoto, S. Magnetovolume Effect in Mn $_3$ Cu $_{1-x}$ Ge $_x$ N Related to the Magnetic Structure: Neutron Powder Diffraction Measurements. *Phys. Rev. B* **2008**, *77* (2), 020409. https://doi.org/10.1103/PhysRevB.77.020409.

(15) Barberon, M.; Fruchart, M. E.; Fruchart, R.; Lorthioir, G.; Madar, R.; Nardin, M. Un nouveau type de deformation orthorhombique dans les perovskites metalliques. *Mater. Res. Bull.* **1972**, *7* (2), 109–118. https://doi.org/10.1016/0025-5408(72)90267-X.

(16) Asano, K.; Koyama, K.; Takenaka, K. Magnetostriction in Mn3CuN. *Appl. Phys. Lett.* **2008**, *92* (16), 161909. https://doi.org/10.1063/1.2917472.

(17) Takenaka, K.; Shibayama, T.; Kasugai, D.; Shimizu, T. Giant Field-Induced Distortion in Mn$_3$ SbN at Room Temperature. *Jpn. J. Appl. Phys.* **2012**, *51* (4R), 043001. https://doi.org/10.1143/JJAP.51.043001.

(18) Takenaka, K.; Ichigo, M.; Hamada, T.; Ozawa, A.; Shibayama, T.; Inagaki, T.; Asano, K. Magnetovolume Effects in Manganese Nitrides with Antiperovskite Structure. *Sci. Technol. Adv. Mater.* **2014**, *15* (1), 015009. https://doi.org/10.1088/1468-6996/15/1/015009.

(19) Boller, H. Komplexcarbide und-nitride mit aufgefülltem U3Si-Typ. *Monatshefte Für Chem.* **1968**, *99* (6), 2444–2449. https://doi.org/10.1007/BF01154362.

(20) Kasugai, D.; Ozawa, A.; Inagaki, T.; Takenaka, K. Effects of Nitrogen Deficiency on the Magnetostructural Properties of Antiperovskite Manganese Nitrides. *J. Appl. Phys.* **2012**, *111* (7), 07E314. https://doi.org/10.1063/1.3672243.

(21) Bang, H.-W.; Yoo, W.; Lee, K.; Lee, Y. H.; Jung, M.-H. Magnetic and Structural Phase Transitions by Annealing in Tetragonal and Cubic Mn3Ga Thin Films. *J. Alloys Compd.* **2021**, *869*, 159346. https://doi.org/10.1016/j.jallcom.2021.159346.

(22) Rodríguez-Carvajal, J. Recent Advances in Magnetic Structure Determination by Neutron Powder Diffraction. *Phys. B Condens. Matter* **1993**, *192* (1–2), 55–69. https://doi.org/10.1016/0921-4526(93)90108-I.

(23) Rodriguez-Carvajal, J.; Gonzalez-Platas, J.; Katcho, N. A. Magnetic Structure Determination and Refinement Using *FullProf*. *Acta Crystallogr. Sect. B Struct. Sci. Cryst. Eng. Mater.* **2025**, *81* (3), 302–317. https://doi.org/10.1107/S2052520625003944.

(24) Pomjakushin, V. On the Magnetic and Crystal Structures of NiO and MnO. *Acta Crystallogr. Sect. B Struct. Sci. Cryst. Eng. Mater.* **2024**, *80* (5), 385–392. https://doi.org/10.1107/S205252062400756X.

(25) Wills, A. S. A New Protocol for the Determination of Magnetic Structures Using Simulated Annealing and Representational Analysis (SARAh). *Phys. B Condens. Matter* **2000**, *276–278*, 680–681. https://doi.org/10.1016/S0921-4526(99)01722-6.

(26) Sharan, A.; Lany, S. Computational Discovery of Stable and Metastable Ternary Oxynitrides. *J. Chem. Phys.* **2021**, *154* (23), 234706. https://doi.org/10.1063/5.0050356.

(27) Rom, C. L.; Smaha, R. W.; Melamed, C. L.; Schnepf, R. R.; Heinselman, K. N.; Mangum, J. S.; Lee, S.-J.; Lany, S.; Schelhas, L. T.; Greenaway, A. L.; Neilson, J. R.; Bauers, S. R.; Tamboli, A. C.; Andrew, J. S. Combinatorial Synthesis of Cation-Disordered Manganese Tin

Nitride MnSnN$_2$ Thin Films with Magnetic and Semiconducting Properties. *Chem. Mater.* **2023**, *35* (7), 2936–2946. https://doi.org/10.1021/acs.chemmater.2c03826.

(28) Wallace, S. K.; Frost, J. M.; Walsh, A. Atomistic Insights into the Order–Disorder Transition in Cu$_2$ZnSnS$_4$ Solar Cells from Monte Carlo Simulations. *J. Mater. Chem. A* **2019**, *7* (1), 312–321. https://doi.org/10.1039/C8TA04812F.

(29) Perdew, J. P.; Burke, K.; Ernzerhof, M. Generalized Gradient Approximation Made Simple. *Phys. Rev. Lett.* **1996**, *77* (18), 3865–3868. https://doi.org/10.1103/PhysRevLett.77.3865.

(30) Blöchl, P. E. Projector Augmented-Wave Method. *Phys. Rev. B* **1994**, *50* (24), 17953–17979. https://doi.org/10.1103/PhysRevB.50.17953.

(31) Kresse, G.; Furthmüller, J. Efficient Iterative Schemes for *Ab Initio* Total-Energy Calculations Using a Plane-Wave Basis Set. *Phys. Rev. B* **1996**, *54* (16), 11169–11186. https://doi.org/10.1103/PhysRevB.54.11169.

(32) Peng, H.; Scanlon, D. O.; Stevanovic, V.; Vidal, J.; Watson, G. W.; Lany, S. Convergence of Density and Hybrid Functional Defect Calculations for Compound Semiconductors. *Phys. Rev. B* **2013**, *88* (11), 115201. https://doi.org/10.1103/PhysRevB.88.115201.

(33) Sun, J.; Ruzsinszky, A.; Perdew, J. P. Strongly Constrained and Appropriately Normed Semilocal Density Functional. *Phys. Rev. Lett.* **2015**, *115* (3), 036402. https://doi.org/10.1103/PhysRevLett.115.036402.

(34) Stoner, E. C.; Wohlfarth, E. P. A Mechanism of Magnetic Hysteresis in Heterogeneous Alloys. *Philos. Trans. R. Soc. Lond. Ser. Math. Phys. Sci.* **1948**, *240* (826), 599–642. https://doi.org/10.1098/rsta.1948.0007.